\documentstyle[prl,aps]{revtex}
\def\be{\begin{equation}}
\def\ee{\end{equation}}
\def\bea{\begin{eqnarray}}
\def\eea{\end{eqnarray}}

\font\trm=cmr10

\font\nrm=cmr9
\font\sbf=cmbx7
\font\latin=cmsy10
\def\npi{\hbox{\latin N$_\pi$}}

\def\gutz{\hbox{\latin P$_{\rm G}$}}
\def\nvpi{\hbox{\latin N$_\varpi$}}

\begin{document}
\draft
\twocolumn[\hsize\textwidth\columnwidth\hsize\csname@twocolumnfalse%
\endcsname

\title{{\hfill\trm SU-ITP 97/16, cond-mat/9705282\bigskip\\}}



\maketitle


]

\noindent{\bf Greiter Replies:}\ \
1.\
In a comment on my recent manuscript\cite{ma} on the $\pi$-particle,
Demler, Zhang, Meix\-ner, and Hanke\cite{com} have raised two points;
I have reservations about one of them, and believe the other to
be valid and excellent.  The point I have reservations about 
concerns the question whether there is a renormalization
of the chemical potential entering eqn.\ (3) of\cite{eugene}
or not; the point I will be happy to elaborate concerns the issue of how
to interpret the low energy resonance peak observed by Meixner
{\it et al.\ }\cite{meix} in the $\pi$-$\pi$ correlation function 
$\pi_d^+(\omega )$
at $n=0.6$ shown in figures 1(a) and 1(b) of their manuscript.  
Those readers who already share my reservations towards 
a possible renormalization of the chemical potential 
may immediately proceed to paragraph 5 below.

\smallskip
2.\
My reply to the first point raised by Demler {\it et al.}\cite{com} is:
The chemical potential in eqns.\ (2) and (3) of\cite{eugene} 
(or (3), (4) and (7) of\cite{ma})
does not get renormalized by the Hubbard interaction U as it is the fully
renormalized, exact and physical chemical potential to begin with.
No self consistency requirement within any applicable approximation 
requires one to assert the contrary; such an assertion is simply incorrect.

To see this, let us go back and understand why the chemical potential
enters into the calculation at hand to begin with.  
If we work in the canonical ensemble,
\be
\widetilde H = 
- t\sum_{\langle ij\rangle\, \sigma} c^\dagger_{i\sigma}c_{j\sigma}
+ {J\over 2}\sum_{\langle ij\rangle} {\bf S}_i\cdot{\bf S}_j 
+ U \sum_i n_{i\uparrow}n_{i\downarrow}, 
\label{eq:hcan}
\ee
and assume $\pi_d^\dagger$
was an approximate eigenoperator of $\widetilde H$, the commutator
\be
\lbrack \widetilde H,\pi_d^\dagger\,\rbrack 
\approx 2Un_\downarrow \pi_d^\dagger \,+\,\hbox{term linear in $J$} 
\equiv \widetilde\omega \pi_d^\dagger
\label{eq:picom}
\ee
would yield the energy of the $N$ particle excited state 
\be
\vert\Psi_\pi^N\rangle = 
\npi \pi_d^\dagger \vert\Psi_0^{N-2}\rangle
\label{eq:psipi}
\ee
relative to the $N-2$ particle ground state $\vert\Psi_0^{N-2}\rangle$:
\be
\widetilde\omega = E_\pi^N -E_0^{N-2}.
\label{eq:omcan}
\ee
The energy of the resonance, however, has to be measured relative to the
ground state containing the same number of particles; the energy
of the $\pi$-particle is therefore given by
\be
\omega_0 = E_\pi^N -E_0^N =\widetilde\omega -2\mu,
\label{eq:omzero}
\ee
where 
\be
\mu = {1\over 2}(E_0^N -E_0^{N-2})
\label{eq:mu}
\ee
is the chemical potential of the fully interacting system.  The formalism
takes care of this correction automatically if we work in the
grand-canonical ensemble and introduce a chemical potential term into
the Hamiltonian, which then fixes the number of particles;
the above discussion merely illustrates that this chemical potential,
independently of how $\lbrack H,\pi_d^\dagger\rbrack $ is evaluated, 
has to be the fully renormalized one.

3.\
Since the question in dispute appears to be not only whether the chemical
potential gets renormalized or not, but whether the energy expectation
value of the $\pi$-particle (\ref{eq:psipi}) is (in the large $U$ limit)
of order $U$ or not, I will address this question from a slightly 
different angle as well.

The energy expectation value of $\pi$-excitation is given by
\be
\omega_0 = 
\langle\Psi_\pi^N\vert \widetilde H \vert\Psi_\pi^N\rangle -
\langle\Psi_0^N\vert \widetilde H \vert\Psi_0^N\rangle,
\label{eq:omdef}
\ee
where $\widetilde H$ and $\vert\Psi_\pi^N\rangle$ are as defined
in (\ref{eq:hcan})
and (\ref{eq:psipi}).  Note that there is no need to
introduce a chemical potential,
and that no approximation has been made.  
It has been demonstrated in paragraph 3 of\cite{ma}
that the state $\vert\Psi_\pi^N\rangle$
has a large amplitude to contain one or two doubly occupied sites,
while the amplitude to find doubly occupied sites in the ground state
$\vert\Psi_0^N\rangle$ vanishes in the large $U$ limit 
of the less-than-half filled Hubbard model.  
Accordingly, the contribution of the Hubbard interaction
$U$ to $\omega_0$ is approximately $2Un_{\downarrow}$, while all other
contributions to $\omega_0$ remain finite as $U\to\infty$. 
Consequently, $\omega_0$ is of order $U$. 
This result can be confirmed numerically by
evaluating (\ref{eq:omdef}) exactly for a finite-size Hubbard cluster.

4.\
In the classic papers on the ground-state energy of a many fermion system
by Kohn, Luttinger, and Ward\cite{lutt} cited in\cite{com},
the small parameter in the perturbation expansion controls the mutual
interactions between all the electrons in the system; one 
starts with a non-interacting Fermi liquid and gradually switches on the
interactions.  The latter will affect the ground state energy 
per particle, and hence renormalize the chemical potential.
(Kohn {\it et al.}\cite{lutt} have demonstrated that this renormalization
can be carried out order by order in perturbation theory.)

In the present case, however, only 
the mutual interaction between the two electrons
added by the $\pi_d^\dagger$ operator and interactions between those two and
the remaining electrons enter the
calculation; only these interactions are controlled by the small parameter 
in any perturbative expansion.  These interactions have no influence on
the ground state energy or the chemical potential.

\smallskip
5.\
I will now turn to the interpretation of the low-energy resonances
observed by Meixner {\it et al.}\cite{meix} at $n=0.6$, a density at 
which the $d$-wave superconducting correlations are no longer present
(figures 1(a) and 2(a) of their manuscript). 

The fact that the $\pi_d^\dagger$ operator is not an approximate eigenoperator 
of the Hubbard model does not imply that its projection
onto a specific subspace cannot be an approximate eigenoperator.
In particular, the problem with the doubly occupied sites exposed in 
paragraphs 2 and 3 of\cite{ma} can be circumvented if 
one sandwiches the $\pi_d^\dagger$ operator
\be
\pi_d^\dagger =\sum_{\hbox{\sbf k}} (\cos k_x -\cos k_y) 
c^\dagger_{\hbox{\sbf k}+\hbox{\sbf Q}\uparrow } 
c^\dagger_{-\hbox{\sbf k}\uparrow }
\ \ \ \ {\bf Q}\equiv (\pi,\pi).
\label{eq:pi}
\ee
in-between two Gutzwiller projectors, 
\be
\gutz \equiv \prod_i (1-n_{i\uparrow} n_{i\downarrow}),
\label{eq:gutz}
\ee
and considers the resulting operator\cite{varpi} 
\be
\varpi_d^\dagger =\gutz \pi_d^\dagger \gutz, 
\label{eq:varpi}
\ee
as a candidate for an approximate eigenoperator of the $t$-$J$ model,
which is obtained from the Hubbard model (\ref{eq:hcan}) by taking
the limit $U\to\infty$.  
Since the $\varpi$-excitation
\be
\vert\Psi_\varpi^N\rangle = 
\nvpi \varpi_d^\dagger \vert\Psi_0^{N-2}\rangle
\label{eq:psivpi}
\ee
has a finite overlap with the $\pi$-excitation (\ref{eq:psipi})
in the thermodynamic limit, I conjecture that the intermediate state
(\ref{eq:psivpi}) is responsible for the sharp resonance peak observed by
Meixner {\it et al.}\cite{meix} in the $\pi$-$\pi$ correlation function
$\pi_d^+(\omega )$\cite{pdef}
for $n=0.6$. 

\smallskip
6.\
The $\varpi$-excitation introduced above, 
however, has a number of problems:

(a)
Since the Gutzwiller projector $\gutz$ does not commute with the
hopping term in (\ref{eq:hcan}),
$\vert\Psi_\varpi^N\rangle$ is not an eigenstate of 
the kinetic part of the $t$-$J$ (or Hubbard) Hamiltonian.

(b)
The $\varpi$ operators do not satisfy all the commutation
relations of the SO(5) algebra\cite{sofive}.  
They can, however, 
be used to rotate an antiferromagnetic order operator into a
$d$-wave superconducting order operator; since
$\gutz$ 
commutes with the spin density wave operator $S_{\hbox{\sbf Q}}^+$\cite{pdef} 
the commutator (17) of\cite{ma} reduces to 
\be
{1\over 2} \lbrack \varpi_d ,S_{\hbox{\sbf Q}}^+ \rbrack
=\gutz \Delta_d \gutz.
\label{eq:dgutz}
\ee
Since the $t$-$J$ Hamiltonian does not allow for doubly occupied sites,
the Gutzwiller projectors in (\ref{eq:dgutz}) have no effect on 
the superconducting order parameter 
$\langle\Psi_0^{N-2}\vert \Delta_d \vert\Psi_0^N\rangle $.

(c)\
Unpublished finite-size studies
by Meixner and Han\-ke\cite{priv}
show a sharp peak in the dynamical spin-spin correlation functions
$\chi_{\hbox{\sbf Q}}^+(\omega )$\cite{pdef}
at $n=0.8$, which occurs at a slightly lower energy than the resonance
in $\pi_d^+(\omega )$ at $n=0.6$ 
shown in figures 1(a) and 2(a) of\cite{meix}.
(The intermediate states in both correlation functions contain the
same number of particles.)
This result constitutes significant evidence against
an interpretation of the
magnetic resonance peak observed in superconducting YBa$_2$Cu$_3$O$_7$
in terms of the $\varpi$-excitation, 
or a resonance in the particle-particle channel in general.

(d)\
One of the side effects of the on-site Hubbard or $t$-$J$ model descriptions
of the CuO-planes in high-T$_{\hbox{\sbf c}}$ superconductors is that
the antiferromagnetic exchange interaction yields
an effective attraction between holes on nearest-neighbor sites,
which is of order $J\langle {\bf S}_i\cdot{\bf S}_{i+1} \rangle$.  
In the CuO-planes,
this attraction is over-compensated by the Coulomb repulsion between
the holes; depending on the details of the screening by the highly 
polarizable O-atoms in between the Cu-atoms, this repulsion is 
by a factor of at least 10 larger than the magnetic attraction.

I conjecture that most of the spectral weight in the low-energy
resonance peaks observed by Meixner {\it et al.}\cite{meix}
in the $\pi$-$\pi$ correlation function $\pi_d^+(\omega )$
will disappear at all densities
once a nearest neighbor repulsion of order $J=4t^2/U$
is introduced to compensate for this ``un-physical'' magnetic attraction.

\smallskip
This work was supported through NSF grant No.\ DMR-95-21888.
Additional support was provided by the NSF MERSEC Program through 
the Center for Materials Research at Stanford University.

\vspace{0.5cm}
\noindent\hbox{Martin Greiter\hfill}

\medskip
\vbox{
\obeylines\nrm 
Department of Physics
Stanford University
Stanford, CA 94305}

\vspace{.5cm}
\noindent\hbox{\nrm May 27, 1997\hfill}

\noindent\hbox{\nrm PACS numbers: 74.72.Bk, 74.25.Ha, 75.40.Gb\hfill}


\vspace{-0.5cm}
\begin{references}
\vspace{-1.6cm}

\bibitem{ma}
M.\ Greiter, {\it Is the $\pi$-particle responsible for the 41 meV
peak in YBa$_2$Cu$_3$O$_7$?}, SU-ITP 97/15, cond-mat/9705049,
submitted to {\em Phys.\ Rev.\ Lett.}.

\bibitem{com}
E.\ Demler, S.C.\ Zhang, S.\ Meixner, and W.\ Hanke, 
{\it Response to Greiter's comment}, cond-mat/9705191.

\bibitem{eugene}
E.\ Demler and S.C.\ Zhang, {\em Phys.\ Rev.\ Lett.\ }{\bf 77}, 4126 (1995). 

\bibitem{meix}
S.\ Meixner, W.\ Hanke, E.\ Demler, and S.C.\ Zhang, {\it Finite-Size
Studies on the SO(5) Symmetry of the Hubbard Model}, cond-mat/9701217,
submitted to {\em Phys.\ Rev.\ Lett.}

\bibitem{lutt}
W.\ Kohn and J.M.\ Luttinger, {\em Phys.\ Rev.\ }{\bf 118}, 41 (1960);
J.M.\ Luttinger and J.C.\ Ward, {\em Phys.\ Rev.\ }{\bf 118}, 1417 (1960).

\bibitem{varpi}
$\varpi$ is generated with $\backslash${\tt varpi} in \TeX.

\bibitem{pdef}
$\pi_d^+(\omega )$ and $\chi_{\hbox{\sbf Q}}^+(\omega )$ are
defined in (2) and (5) of\cite{meix}, or (8) and (9) of\cite{ma};
$S_{\hbox{\sbf Q}}^+$ is defined in (3) of\cite{meix} or (10) of\cite{ma}.

\bibitem{sofive}
S.C.\ Zhang, {\em Science\ }{\bf 275}, 1089 (1997). 

\bibitem{priv}
S.\ Meixner, private communication.

\end{references}
\end{document}